\documentclass[aps,prd,noshowpacs,preprintnumbers,nofootinbib,floatfix,floats,groupedaddress,twocolumn]{revtex4}
\usepackage{bm}
\usepackage{latexsym}
\usepackage{dcolumn}
\usepackage{amsmath,amsfonts,amssymb}
\usepackage{graphicx,epsfig}
\usepackage{color}

\def\eq#1{{Eq.~(\ref{#1})}}

\def\ll{Lanczos-Lovelock Lagrangian}
\def\ee{Einstein-Hilbert Lagrangian}
\def\e{{\cal G}}
\def\ea{Einstein-Hilbert action}
\def\fe{field equations}
\def\bg{bulk $\Gamma \Gamma$ term}
\def\hr{holographic relationship}

\def\eq#1{{Eq.~(\ref{#1})}}

\def\eq#1{{Eq.~(\ref{#1})}}

\newcommand{\LL}{Lanczos-Lovelock}

\reversemarginpar

\begin{document}

\title{Holography in Action}

 \author{Sanved Kolekar}
 \email{sanved@iucaa.ernet.in}
 \author{T.~Padmanabhan}
 \email{paddy@iucaa.ernet.in}
 \affiliation{IUCAA,
 Post Bag 4, Ganeshkhind, Pune - 411 007, India}

\date{\today}
\begin{abstract}
Einstein-Hilbert action and its natural generalizations to higher dimensions (like the \LL\ action) have certain peculiar features. All of them  can be separated into a bulk  and a surface term, with a specific (``holographic") relationship between the two, so that either term can be used to extract information about the other. Further, the surface term leads to entropy of the horizons on-shell. It has been argued in the past that these features are impossible to understand in the conventional approach but find a natural explanation if we consider gravity as an emergent phenomenon. We provide further support for this point of view in this paper. We describe an alternative decomposition of the \ea\ and \LL\ action into a new pair of surface and bulk terms, such that the surface term becomes Wald entropy on a horizon and the bulk term is the energy density (which is the ADM Hamiltonian density for Einstein gravity). We show that this new pair also obeys a \hr\ and give a thermodynamic interpretation to this relation in this context.  Since the bulk and surface terms, in this decomposition, are related to energy and entropy, the holographic condition can be thought of as analogous to inverting the expression for entropy given as a function of energy $S=S(E,V)$ to obtain the  energy $E=E(S,V)$ in terms of the entropy in a normal thermodynamic system. Thus the holographic nature of the action allows us to relate the descriptions of the same system in terms of two different thermodynamic potentials.  Some further possible generalizations and  implications are discussed.

\end{abstract}

\maketitle
\vskip 0.5 in
\noindent
\maketitle

\section{Motivating the holographic actions for gravity} \label{sec:intro}

The peculiar relationship between thermodynamics and dynamics of horizons, known for four decades \cite{oldies},  is now being slowly recognized as indicating a more fundamental principle in which gravity can be viewed as an emergent phenomenon like fluid mechanics or elasticity. (For a recent review of this approach, see \cite{rop}; for a sample of papers,  implementing and discussing this paradigm in different ways, see ref. \cite{others}.) Such a point of view draws support from several pieces of evidence of which we may mention the following: 

(a) The  field equations of gravity reduce to a thermodynamic identity on the horizons in a wide variety of models much more general than just Einstein's gravity \cite{sphcqg, ronggen}. As pointed out first in \cite{surfaceaction}, and confirmed by several pieces of later work, the thermodynamic paradigm seems to be applicable to a wide class of theories much more general than Einstein gravity in 4-dimensions.
 
 (b) It is possible to obtain the field equations of gravity --- again for a wide class of theories --- from purely thermodynamic considerations (see e.g. \cite{reviews2}). 
 
 (c) One can obtain an equipartition law analogous to $E=(1/2)n k_BT$ for the density $n$ of microscopic degrees of freedom in any static geometry \cite{equi} providing a direct window to microscopic physics in the thermodynamic limit.
 
In such an approach, geometrical variables like the metric etc are derived concepts (like e.g., pressure, density etc. of a gas) and the dynamical equations governing them can be derived from  the thermodynamic limit of an underlying microstructure, say, by extremising a suitably defined entropy functional \cite{reviews2}.  But, on the other hand, 
we know from standard textbook description, that one can obtain the field equations for gravity from a  action functional in which the metric is varied. This raises the question: 

\textit{If gravity is indeed an emergent phenomenon, should not the conventional action functionals contain some signature of this fact ?}

After all, field equations ``know" that there exist an alternative, emergent, interpretation for the dynamics. Hence it seems reasonable to assume that this information must be embedded in the action functionals describing theories of gravity in some manner. There is evidence that this is indeed the case \cite{grf2002}. 
 There are peculiar holographic relations between the surface and bulk terms in the action functionals describing several theories of gravity and the surface term in the action is closely related to entropy of horizons  in all these theories. Our aim is to elucidate this further.

Let us begin by reviewing some known facts and interpret them in a manner useful for our discussion, starting from \ea. It is well-known that 
the \ea\ can be separated into a bulk term and a surface term. (Several facts related to this decomposition was known fairly early in the literature and in particular to Einstein himself \cite{gammagamma}; for a modern textbook description, see chapter 6 of ref. \cite{gravitation}.) The bulk term (the `Gamma-Gamma' term) depends on the metric and its first derivatives and is quadratic in the latter;  the surface term arises from integrating a total divergence and contains both the normal and tangential derivatives of the metric on the boundary. Because of the dependence of the surface term on the normal derivatives of the metric, the action principle cannot be formulated in the usual manner. In general, there are two ways of handling this issue: 

(a) One can add an extra term \cite{gibbonhawking} to the \ea\ such that the variation of this term precisely cancels the unwanted terms arising in the variation of the original surface term. 

(b) One can simply ignore the surface term in the \ea\ and vary the bulk term keeping the metric fixed at the boundary; even though the bulk term is not generally covariant, the resulting field equations are indeed covariant. 

In either approach, it is \textit{only} the  variations of the \bg\ that contribute to the \fe. That is, the field equations (and their solutions) do \textit{not} depend in any way on the surface term.
It is therefore a mystery --- in the conventional approach --- that the surface term, which is ignored while obtaining the field equations can be used to determine the entropy of the horizons that arise in the theory! 

The solution to this mystery was first pointed out in \cite{grf2002} where it was emphasized that the bulk and surface terms in Einstein-Hilbert action are connected by a peculiar relation:
\begin{equation}
L_{\rm sur}=-\partial_i\left(g_{ab}\frac{\partial L_{\rm bulk}}{\partial (\partial_ig_{ab})}\right)
\label{holoeh1}
\end{equation}
which allows the information about either one to be extracted from the other. It was also shown that one can obtain the bulk action from the surface term if one adopts the thermodynamic perspective of gravity. Later on, these ideas were generalized to a wide class of models \cite{ayantp} including the Lanczos-Lovelock \cite{LL} models. The relationship between the bulk and boundary terms in the action was termed `holographic' because the information about the bulk action functional (from varying of which we can obtain the dynamical equations) is encoded in the boundary action functional. In this paper, we shall continue to use this terminology `holographic action' with this understanding.

The holographic nature of the action fits very well with the thermodynamic approach to gravity and can be thought of the hidden signal in the action functionals indicating that the description of gravity is an emergent one.
In fact one can provide very general arguments to suggest that the action functional describing \textit{any} theory of gravity that obeys the principle of equivalence and principle of general covariance will have a bulk and boundary term related holographically (see e.g.,\cite{reviews2}.)
If this is the case, one would like to explore this connection further and see what insights it can provide. 
In particular, we would like to address the following concrete questions: 

(a)
Of the two terms --- bulk and boundary --- the boundary term has a clear interpretation as being related to the entropy of horizons. But the physical interpretation of the bulk term is unclear and one would like to have a thermodynamic interpretation for the same. 

(b) One would like to know whether there is something special in the particular  decomposition of the \ea\ so that it admits a holographic relationship. Or can the holographic relationship arise in other contexts when we decompose the \ea\ into a bulk and  surface term in a different manner ?

It turns out that the answers to these two questions are closely related. We will show that there is an alternative way of decomposing the action functionals in the case of static geometries which gives simple thermodynamic interpretation to both bulk and boundary terms as energy and entropy. More importantly this decomposition is also holographic so that one can extract the information about the energy from entropy and vice-versa.

The paper is organized as follows. In section II, we briefly review the previous work done  as regards  holography of action. We also setup the notations that we use in the rest of the paper. In section III, we look at the decomposition of the \ea\ into a new pair of surface and bulk term, different from the usual splitting. We show this new pair also obeys a \hr. We will study this decomposition from a thermodynamic point of view and give a meaning to the holographic relationship as playing the same role as the relationship between two thermodynamic potentials --- viz., the entropy and the energy of a  thermodynamic system. In section IV, we generalize the results of section III to the Lanczos-Lovelock models of gravity. Further we prove a general result which gives a \hr\ between any decomposition of the Lanczos-Lovelock action into a arbitrary bulk and a surface term provided the surface term is homogeneous in its dynamical degrees of freedom. The conclusions are discussed in section V.

The metric signature is $(-, +, +, \ldots ,+)$, and all the fundamental constants such as $G, \hbar$ and $c$ have been set to unity. Latin indices run from 0-3, whereas Greek indices run from 1-3.

\section{Holography of the gravitational action}
We begin by expressing the Einstein-Hilbert Lagrangian $L_{EH} = R$ in a manner which will be  convenient for our further discussions and generalization to \LL\ models. We write

\begin{equation}
L_{EH}  \equiv  Q_a^{\phantom{a}bcd}R^a_{\phantom{a}bcd}=Q^{ab}_{cd}R^{ab}_{cd}\equiv \delta^{cd}_{ab}R^{ab}_{cd} 
\label{leh} 
=  R 
\end{equation}
where
\begin{equation}
 Q^{ab}_{cd}=\delta^{cd}_{ab}=\frac{1}{2}(\delta^c_a\delta^d_b-\delta^d_a\delta^c_b)
\end{equation}
Here $\delta_{ab}^{cd}$ is the alternating (`determinant') tensor. The tensor  $Q^{abcd}$ is the only fourth rank tensor that can be constructed from the metric (alone) that has 
\begin{itemize}
\item[ ] 
(i) all the symmetries of the curvature tensor 
\item[ ]
(ii) zero divergence on all indices, $\nabla_a Q^{abcd}=0$ etc.
\end{itemize}
The total action written as a sum of the \ea\ and the matter action $A_m$ so that:
 \begin{equation}
A_{total} = \int_\mathcal{V} d^Dx\sqrt{-g}\, L_{EH} + \int_\mathcal{V} d^Dx\sqrt{-g}\, L_{matter}
\end{equation}  
The variation of the metric in this action, after ignoring the surface terms, leads to the Einstein field equations
\begin{eqnarray}
G_{ab} &= & Q_a^{\ cde} R_{bcde}-\frac{1}{2}g_{ab}R \nonumber \\
& = & R_{ab}-\frac{1}{2}g_{ab}R =\frac{1}{2}  T_{ab}
\label{fieq}
\end{eqnarray}
where $T_{ab}$ is the stress-energy tensor obtained from the variation of the matter part of the total action.

We will now state a feature of the \ea\ relevant to our discussion, viz., its decomposition into a bulk term and a surface term. It can be shown \cite{ayantp} that when the Lagrangian has the form $Q_a^{\phantom{a}bcd}R^a_{\phantom{a}bcd}$ with   $Q^{abcd}$ obeying the two properties (i) and (ii) mentioned in the last paragraph, there is natural decomposition of the Lagrangian into a bulk and surface terms which are holographically related. (The actual form of $Q^{abcd}$ is irrelevant as long as it has the symmetries of the curvature tensor and is divergence-free.). In the case of \ea\ the $\sqrt{-g}L_{EH}$  can be written as a sum $L_{bulk}+L_{sur}$ where $L_{bulk}$ is quadratic in the first derivatives of the metric and $ L_{sur}$ is a total derivative which leads to a surface term in the action: 
 \begin{eqnarray}
\sqrt{-g}L_{\rm EH} && =2\partial_c\left[\sqrt{-g}Q_a^{\phantom{a}bcd}\Gamma^a_{bd}\right]
+2\sqrt{-g}Q_a^{\phantom{a}bcd}\Gamma^a_{dk}\Gamma^k_{bc} \nonumber \\ \nonumber \\
&& \equiv L_{\rm sur} + L_{\rm bulk} 
\label{stdsplit}
 \end{eqnarray} 
with
\begin{eqnarray}
L_{\rm bulk} && =2\sqrt{-g}Q_a^{\phantom{a}bcd}\Gamma^a_{dk}\Gamma^k_{bc}; \label{lb} \\ \nonumber \\
L_{\rm sur} && =2\partial_c\left[\sqrt{-g}Q_a^{\phantom{a}bcd}\Gamma^a_{bd}\right] \label{ls}
\end{eqnarray}
 It is well known that one can obtain the Einstein's equations, \eq{fieq}, by varying only $L_{bulk}$  keeping $g_{ab}$ fixed at the boundary (see Appendix [\ref{appggvary}] for a brief demonstration).
What is more remarkable is the fact that there exists a   simple relation between $L_{\rm bulk}$ and $L_{\rm sur}$ allowing $L_{\rm sur}$ to be determined completely by $L_{\rm bulk}$
\cite{grf2002, holo}. 
It is given as
\begin{equation}
L_{\rm sur}=-\frac{1}{[(D/2)-1]}\partial_i\left(g_{ab}\frac{\partial L_{\rm bulk}}{\partial (\partial_ig_{ab})}\right)
\label{holoeh}
\end{equation}
which is a generalization of \eq{holoeh1} to $D$-dimensions.
As discussed in Sec. \ref{sec:intro}, we call such a relation  holographic.

All  the above results generalize to a class of actions known as the Lanczos-Lovelock action \cite{LL} which is a generalization of the Einstein-Hilbert action. The \ll\ is constructed as a special product of $m$ curvature tensors $R^{ab}_{cd}$ given by
\begin{equation}
L_m=\delta^{1357...2k-1}_{2468...2k}R^{24}_{13}R^{68}_{57}
....R^{2k-2\,2k}_{2k-3\,2k-1}; \qquad k=2m
\end{equation} 
where $k=2m$ is an even number. For $m=1$, the $L_m$ reduces to the \ee\ in the form given in \eq{leh}. The $L_m$ is clearly a homogeneous function of degree $m$ in
the curvature tensor $R^{ab}_{cd}$ so that it  can also  be expressed in the form:
\begin{eqnarray}
L &&=\frac{1}{m}\left(\frac{\partial L}{\partial R^a_{\phantom{a}bcd} }\right)R^a_{\phantom{a}bcd}\equiv \frac{1}{m}P_a^{\phantom{a}bcd}R^a_{\phantom{a}bcd} \nonumber \\
&& = Q_a^{\phantom{a}bcd}R^a_{\phantom{a}bcd}
\end{eqnarray} 
where we have defined $P_a^{\phantom{a}bcd}\equiv (\partial L/\partial R^a_{\phantom{a}bcd} )$
so that $P^{abcd}=mQ^{abcd}$. Hence $Q^{abcd}$ inherits  all the symmetries of the curvature tensor. It can be also directly verified that for these Lagrangians 
\begin{equation}
\nabla_cP^{ijcd}=m\nabla_cQ^{ijcd}=0
\label{condition}
\end{equation} 
Because of the symmetries,  $Q^{abcd}$ is divergence-free in \textit{all} indices. So the $Q^{abcd}$ satisfies the two conditions (i) and (ii) mentioned at the beginning of this section. 

The  total action  is obtained from adding the \ll\ to the matter Lagrangian and integrating over a $D$ dimensional region $\mathcal{V}$. The variation of this action, ignoring the boundary terms on $\partial\mathcal{V}$ leads to the following \fe
\begin{eqnarray}
{\cal G}_{ab} & \equiv & P_a^{\ cde}R_{bcde}-\frac{1}{2}g_{ab}L \nonumber \\
& \equiv & {\cal R}_{ab}-\frac{1}{2}g_{ab}L=\frac{1}{2}  T_{ab}
\label{fiell}
\end{eqnarray}
which, of course, reduces to \eq{fieq} for $m=1$. The notation with calligraphic font is motivated by the fact that ${\cal G}_{ab}\to G_{ab}$ and ${\cal R}_{ab}\to R_{ab}$ in Einstein's theory.

Since $Q^{abcd}$ also satisfies the same two properties (i) and (ii) here too $L_m$ can be separated \cite{ayantp} in to bulk and surface terms as
\begin{eqnarray}
\sqrt{-g}L_{\rm m} && =2\partial_c\left[\sqrt{-g}Q_a^{\phantom{a}bcd}\Gamma^a_{bd}\right]
+2\sqrt{-g}Q_a^{\phantom{a}bcd}\Gamma^a_{dk}\Gamma^k_{bc} \nonumber \\
&& \equiv L_{\rm sur} + L_{\rm bulk} 
\label{llsum}
 \end{eqnarray} 
One crucial difference between Einstein gravity and the more general \LL\ models is the following: In Einstein gravity $L_{bulk}\simeq Q\Gamma\Gamma$ is quadratic in the first derivatives of the metric and does not involve second derivatives of the metric. This is because in this case, $Q^{abcd}$ depends only on the metric. In the case of ($m>1$) \LL\ models, $Q^{abcd}$ will have a nontrivial dependence on the curvature tensor and hence on the second derivatives of the metric. Therefore, $L_{bulk}$ now depends on both first derivatives of the metric as well as second derivatives and hence, while   
varying the action (based on either $L_m$ or $L_{bulk}$) we need to keep both  the metric \textit{and} its normal derivatives fixed at the boundary to get the field equations. It can be shown that (see Appendix [\ref{appggvary}]) under these variations both $L_m$ and $L_{bulk}$  lead to the same field equations in \eq{fiell}. (In principle, one can add counterterms to the general \LL\ action to make it well-defined \cite{llcounter}; but the nature of these counterterms, in general, is quite complicated. Fortunately, this is irrelevant to our discussion.)

Furthermore, as in the case of \ee,  the $L_{sur}$ of the \ll\  can be obtained from the $L_{bulk}$ as \cite{ayantp}
\begin{equation}
[D/2-m]L_{sur}=-\partial_i\left[ g_{ab}\frac{\delta L_{bulk}}{\delta(\partial_ig_{ab})}+\partial_jg_{ab}\frac{\partial L_{bulk}}{\partial(\partial_i \partial_j g_{ab})} \right]
\label{hololl}
\end{equation}
where differentiation by $\delta$ is the Euler derivative.
This is a natural generalization of the holographic relation of \eq{holoeh} and reduces to \eq{holoeh} for $m=1$.
We also mention that the surface term $A_{sur}$ evaluated on the horizon gives one quarter of the area of the boundary when the boundary is a horizon \cite{holo} in Einstein-Hilbert gravity while it is proportional to the corresponding Wald entropy of the horizon \cite{wald} in the case of Lanczos-Lovelock gravity \cite{ayantp} with the proportionality constant being $1/m$. In the absence of the holographic relation between the surface and bulk terms, this fact is difficult to understand because the field equations are independent of the boundary term. 
The holographic relationship in \eq{holoeh} and \eq{hololl}  explains how the two terms of interrelated thereby offering a possible reason why the surface term might have a physical meaning on-shell. We shall now probe these aspects further.

\section{An alternative look at Einstein-Hilbert Action}

We are interested in providing a thermodynamic interpretation to the action functionals in the theories of gravity, taking a clue from the fact the surface term is related to horizon entropy. Since the notion of a temperature and thermal equilibrium is well-defined in static situations, we shall consider the class of all \textit{static} spacetime metrics to elucidate the thermodynamic relationships. 

Our first task is to introduce an alternative decomposition of Einstein-Hilbert action into a surface and bulk term, in any static spacetime.
Using the time-time component of the Einstein tensor $G^a_b$ in \eq{fieq}, the \ee \ can be expressed as
\begin{equation}
L_{EH}=-2G^0_0+2R^0_0
\end{equation}
In any static spacetime, the $R^a_0$ components, in particular $R^0_0$, can be expressed as a divergence term \cite{holo}
\begin{equation}
R^0_0=\frac{1}{\sqrt{-g}}\partial_\alpha(\sqrt{-g}g^{0k}\Gamma^\alpha_{0k})
\label{roo}
\end{equation}
This is most easily seen from noting that any static spacetime has a timelike Killing vector which, for a natural choice for the time coordinate, has the components $\xi^a=(1,\textbf{0})$. The standard identity satisfied by the Killing vector now gives:
\begin{equation}
R^a_j\xi^j=R^a_0=\nabla_b\nabla^a\xi^b=\frac{1}{\sqrt{-g}}\partial_b(\sqrt{-g}\nabla^a\xi^b)
\end{equation}
where the last relation follows from the fact that $\nabla^a\xi^b$ is an antisymmetric tensor.
\eq{roo} now follows directly on noticing that all quantities are time-independent.
Hence, we see that \ee \ for static spacetimes can be expressed as a sum of a bulk term and a surface term in the form:
\begin{equation}
L_{EH}=R=-2G^0_0+2\frac{1}{\sqrt{-g}}\partial_\alpha(\sqrt{-g}g^{0k}\Gamma^\alpha_{0k})
\label{ehstaticsplit}
\end{equation}
Since $G^0_0$ contains second derivatives of the metric, it is obvious that this decomposition is different from the standard decomposition in \eq{stdsplit}.
 We shall now describe several interesting features of this decomposition.

 To begin with, it provides yet another variational principle for obtaining the field equations in the static case. As proved in Appendix [\ref{appg00vary}], the variation of $-2G^0_0$ keeping the static metric fixed on the boundary leads to the usual \fe, viz. the Einstein's equations. 

Second, the action functional as well as the two terms in it has a direct thermodynamic interpretation which is clear in the Euclidean sector obtained by replacing $t$ with $it$. In any static spacetime with a suitable gauge, one has the result:
$-2G^0_0=-16\pi{\cal H}_{ADM}$.
 We can now express the Euclidean Einstein-Hilbert action as an integral
over $R=2(-8\pi{\cal H}_{ADM}+R^0_0)$. Since the spacetime is static, the integrand is independent of time and we limit
the time integration to a finite range $(0,\beta)$ to get a finite result in the Euclidean sector. Converting the volume integral of $R^0_0$ over 3-space
to a surface integral over the 2-dimensional boundary,  we can write \cite{holo} the Euclidean Einstein-Hilbert action in any static spacetime as 
\begin{eqnarray}
A_{E}
&=&-\beta\int N\sqrt{h} d^3x\; {\cal H}_{ADM} \nonumber \\ 
&& +\frac{\beta}{8\pi} \int  d^2x\sqrt{\sigma}\; 
Nn_\alpha(g^{0k}\Gamma^\alpha_{0k})\nonumber\\
&\equiv&-\beta E + S
\label{two}
\end{eqnarray}
where  $N=\sqrt{g^E_{00}}$ is the lapse function, $h$ is the determinant of the spatial metric
and $\sigma$ is the determinant of the 2-metric on the surface. (We have also put in a factor of $1/16\pi$ such that $L_{EH} = (1/16\pi) R$ ).
In a class of  static metrics with a horizon and associated temperature, the time interval has natural 
 periodicity in $\beta$, which can be identified with the inverse temperature. Once this identification is made, the $\beta N$ factor in Eq.(\ref{two}) 
 is  exactly what is needed to give the local Tolman temperature $T_{loc}=\beta^{-1}_{loc}\equiv(\beta N)^{-1}=T/\sqrt{-g_{00}}$. So
 we are actually integrating $\beta_{loc} {\cal H}_{ADM}$ over all space, as one should, and we take the resulting quantity to be $\beta E$. (One can think of $E$ as the thermally averaged energy, obtained with a weightage factor which is the local inverse temperature.)  When the 2-surface is a horizon,
 the integral over $R^0_0$ gives one quarter of the area of the horizon which is the expression for entropy \cite{holo}. (We will show in the next section that $\mathcal{R}^0_0$ is in fact equal to $\beta Q$ where $Q$ is the Noether charge as used by Wald \cite{wald} to define entropy.) This allows identification of the two terms with energy and entropy; together the euclidean Einstein action can be interpreted as giving the (negative of) free energy of space time.

With this thermodynamic interpretation of the action, one can interpret extremising the integral over $-2G^0_0$ while keeping the surface term fixed at the boundary, 
as extremising the bulk energy of the  of static spacetime while keeping the entropy fixed. From previous work we know that the resulting field equations can be interpreted as the thermodynamic identity $dE=TdS-PdV$ on the horizon. In the usual  thermodynamic systems, if we know the entropy functional $S(E,V)$, we can obtain the other thermodynamic variables like $(T,P)$ etc. of the system. Alternatively, one can invert the form of $S(E,V)$ to get energy $E(S,V)$ in terms of entropy. Similarly, in the case of Einstein-Hilbert action, we can  consider the extremisation of $-2G^0_0$, as equivalent to obtaining the thermodynamic relation $dE=TdS-PdV$ (which is the same as field equations of the theory) from an energy functional. Further, if gravity is a truly a emergent, thermodynamic, phenomenon of an underlying microscopic  theory then one should be able to invert the energy functional $E(S,V)$ of gravity to obtain the $S(E,V)$ functional. This motivates us to ask: Can we obtain the surface term $2R^0_0$ directly from the bulk term $-2G^0_0$?
 
 Holography again answers this question. Even in the new decomposition we have introduced, the two terms continue to be related holographically.  We can show by direct computation that $L_{\rm sur}=2\sqrt{-g}R^0_0$ and $L_{\rm bulk}=-2\sqrt{-g}G^0_0$ are related by:
\begin{equation}
L_{\rm sur}=-\frac{1}{[(D/2)-1]}\partial_i\left(g_{ab}\frac{\delta L_{bulk}}{\delta(\partial_ig_{ab})}+\partial_jg_{ab}\frac{\partial L_{bulk}}{\partial(\partial_i \partial_j g_{ab})}\right)
\label{ehholo}
\end{equation}
(see Appendix[\ref{apphee}] for the proof).
One should note that \eq{ehholo} is the general expression of holographic relation for Einstein-Hilbert action. It reduces to the form in \eq{holoeh} in the standard decomposition because $L_{\rm bulk} = 2\sqrt{-g}Q_a^{\phantom{a}bcd}\Gamma^a_{dk}\Gamma^k_{bc}$ is independent of the second derivatives of the metric.

In this analysis, we have  given a physical meaning to the holographic relationship in gravitational action. It is seen as  playing the same role as the relationship between two thermodynamic potentials, viz. the entropy and the energy of a normal thermodynamic system. These results also suggest that the dynamics of spacetime can also be encoded in an entropy functional at the boundary of the spacetime and one could  obtain the \fe\ from this entropy functional. Thus one could in principle formulate a theory of gravity completely by specifying only these entropy functionals at the boundary. Some recent attempts in this direction has been made in ref.\cite{use}

\section{Generalization to Lanczos-Lovelock Theory}

It has been repeatedly noticed in the literature that the thermodynamic aspects transcend Einstein's theory and  occurs in any reasonable theory of gravity that obeys principle of equivalence and general covariance. If we further demand that the field equations should not be of degree higher than two in the metric, one is led to the \LL\ models. Just like several other thermodynamic features,  the results obtained above are also applicable to \LL\ models.
We shall now describe this generalization
in a manner similar to the discussion in the previous section.

Using the time-time component of \eq{fiell}, the $m$-th order \ll\  can be expressed as
\begin{equation}
L_{m}=-2{\cal G}^0_0+2P^{0cde}R_{0cde}
\label{fstsplit}
\end{equation}
In any stationary spacetime, the $P^{acde}R_{0cde}$ components, in particular $P^{0cde}R_{0cde}$, can be expressed as a divergence term 
\begin{equation}
P^{0cde}R_{0cde}=\frac{1}{\sqrt{-g}}\partial_\alpha(\sqrt{-g}P_a^{\ b\alpha 0}\Gamma^a_{0b})
\label{lroo}
\end{equation}
This is again most easily seen from the identity for the Killing vector $\xi^a=(1,\textbf{0})$ 
\begin{equation}
P^{0cdb}R_{acdb}\xi^a=P^{0cdb}\nabla_c\nabla_d\xi_b=\frac{1}{\sqrt{-g}}\partial_c(\sqrt{-g}P^{0cdb}\nabla_d\xi_b)
\end{equation}
where the last relation follows from the fact that $\nabla_cP^{acdb}=0$ and $P^{acdb}$ is antisymmetric in its first two indices.
\eq{lroo} now follows directly on noticing that all quantities are time-independent.
Hence, the \ll\  for static spacetimes can be expressed as a sum of a bulk term and a surface term
\begin{equation}
L_{m}=-2{\cal G}^0_0+2\frac{1}{\sqrt{-g}}\partial_\alpha(\sqrt{-g}P_a^{\ b\alpha 0}\Gamma^a_{0b})
\label{llsp}
\end{equation}
which is a direct generalization of the result in \eq{ehstaticsplit} for Einstein's theory.

The $L_{sur}=2\sqrt{-g}P^{0cdb}R_{0cdb}$ in \eq{fstsplit} has a nice physical interpretation.
It is actually the Noether charge density (i.e., the time component of Noether current) which arises from the diffeomorphism invariance of the theory.
To see this, we re-derive \eq{llsp} using the definition of Noether charge along the lines of ref.\cite{rop}. 
Consider the variation of a \ll, which can be expressed in the form (see appendix [\ref{appllvary}])
\begin{equation}
\delta(L\sqrt{-g})=\sqrt{-g} \left( \e_{ab}\delta g^{ab}+\nabla_a \delta\upsilon ^a \right)
\label{vary}
\end{equation}
where
\begin{eqnarray}
\e_{ab}=P_a^{\ cde}R_{bcde}-\frac{1}{2}g_{ab}L \\
\delta\upsilon ^a=2P^{cbad}(\nabla_b\delta g_{dc}) 
\end{eqnarray}
and
\begin{equation}
P_a^{\ bcd}=\frac{\partial L}{\partial R^a_{\ bcd}}
\end{equation}
When the variations in $\delta g^{ab}$ arise due to the diffeomorphism $x^a \rightarrow x^a +q ^a$, then we have $\delta g^{ab}=\nabla^a q^b+\nabla^b q^a$ and $\delta(L\sqrt{-g})=-\sqrt{-g}\nabla_a (Lq^a)$. Substituting these in \eq{vary} and using the Bianchi identity $\nabla_a \e^{ab}=0$, we obtain a conservation law $\nabla_a J^a=0$, for the Noether current,
\begin{equation}
J^a=Lq ^a + \delta_q v^a +2\e^{ab}q_b
\end{equation}
where $\delta_q v^a$ is the variation of the surface term when the variation in the metric $\delta g^{ab}$ is due to the diffeomorphism. Since $J^a$ is divergenceless, it is convenient to write $J^a$ as $J^a=\nabla_b J^{ab}$ where $J^{ab}$ is an antisymmetric tensor. For Lanczos-Lovelock Lagrangians, knowing the variation of the surface term $\delta_q v^a$, one can obtain the explicit form of  $J^{ab}$ to be \cite{Deruelle:2003ps,rop}
\begin{equation}
J^{ab}=2P^{abcd}\nabla_cq_d
\label{jab}
\end{equation}
As is well-known, the expression of $J^a,J^{ab}$ etc. are not unique; in what follows we shall use the expressions quoted above.

In the case of static spacetimes with a Killing vector $\xi^b$, it is natural to consider the Noether current corresponding to $q^a=\xi^a$. When $\xi^a$ satisfies the Killing equations at an event $\mathcal{P}$, the variation $\delta_\xi v^a$ vanishes at event $\mathcal{P}$ and we get 
\begin{equation}
J^a=L\xi ^a +2\e^{ab}\xi_b
\label{noecur}
\end{equation}
In particular, when $\xi $ is a timelike Killing vector given as $\xi = (1,\textbf{0})$ for static spacetimes, we get
\begin{equation}
J^0=L + 2\e^0_0
\label{as}
\end{equation}
and \eq{jab} becomes:
\begin{equation}
J^{0c}=2P_a^{\ bco}\Gamma^a_{0b}
\end{equation}
Hence, the Lagrangian for static spacetime can be written as a sum of a bulk term and a surface term.
\begin{equation}
L=-2\e^0_0 + \nabla_a J^{0a}
\end{equation}
which is the same as \eq{llsp} and  allows the identification of the divergence term as the time component of Noether current. (Of course, since Einstein's theory is a special case of \LL\ models, this interpretation of the surface term holds for Einstein gravity as well.) We shall now show that the results obtained in the last section for Einstein's theory continue to hold in the present case.
 
 To begin with let us consider the thermodynamic interpretation of the  euclideanised action \cite{rop}. 
 The conserved Noether current for the displacement $x^a \to x^a + \xi^a$ is given by \eq{noecur} . We will work in the Euclidean sector and integrate this expression
 over a constant-$t$ hypersurface with the measure $d\Sigma_a = \delta_a^0 N\sqrt{h}\, d^{D-1} x$
where $g^{E}_{00} = N^2$ and $h$ is the determinant of the spatial metric. Multiplying by
the period $\beta$ of the imaginary time, we get
\begin{eqnarray}
\beta \int J^a d \Sigma_a 
&=& \beta \int 2\e^a_b \xi^b d\Sigma_a  + \beta \int L \xi^a d \Sigma_a\nonumber\\
&=& \int  (\beta  N) 2\e^a_b u_au^b \sqrt{h}\, d^{D-1}x  \nonumber \\
&& + \int_0^\beta dt_E \int L \sqrt{g}\, d^{D-1}x
\end{eqnarray} 
where we have introduced the four velocity  $u^a = \xi^a/N = N^{-1}\delta^a_0$
of observers moving along the orbits of $\xi^a$ and the relation $d\Sigma_a = u_a\sqrt{h}\, d^{D-1} x$.  The term  involving the Lagrangian gives the Euclidean action
for the theory. In the term involving $2\e_{ab}$ we note that $\beta N \equiv \beta_{\rm loc}$ corresponds to the correct redshifted local temperature.
Hence, taking a cue from our procedure for Einstein-Hilbert action, here too we define the (thermally averaged) energy $E$ as
\begin{eqnarray}
\int  (\beta  N) 2\e^a_b u_au^b \sqrt{h}\, d^{D-1}x &=& \int  \beta_{\rm loc}   2\e^a_b u_au^b \sqrt{h}\, d^{D-1}x \nonumber \\
&\equiv& \beta E
\end{eqnarray} 
We thus get 
\begin{equation}
A_E= \beta \int J^a d \Sigma_a -\beta E
\end{equation} 
The first term involving the Noether charge is just the Wald entropy, which continues to hold true in the Euclidean sector.
Therefore, we find that 
\begin{equation}
A_E= S- \beta E = -\beta F 
\end{equation} 
where $F$ is the free energy.  
Thus we have a thermodynamic interpretation for the Lanczos-Lovelock action in the case of static spacetimes. 

The next question to ask, as in the case of Einstein-Hilbert action, is: Are the bulk term and surface terms related by holography? The answer again is `yes'. One can show by direct --- but somewhat more involved calculation outlined in Appendix [\ref{apphll}] --- that
 $L_{\rm sur}=2\sqrt{-g}P^{0cde}R_{0cde}$ is related to $L_{\rm bulk}=-2\sqrt{-g}\e^0_0$ through the holographic relation:
\begin{equation}
[D/2-m]L_{sur}=-\partial_i\left[ g_{ab}\frac{\delta L_{bulk}}{\delta(\partial_ig_{ab})}+\partial_jg_{ab}\frac{\partial L_{bulk}}{\partial(\partial_i \partial_j g_{ab})} \right] \\
\label{mainresult}
\end{equation}
 This is the same relation as \eq{hololl}.
We thus find that the holographic relation between the surface term and the bulk term is not only valid for \ee \ but also to \ll\  which shares the basic geometric structure of \ee.\ Further we see that the relation is true for the Lagrangian written in two different ways (see \eq{llsp} and \eq{llsum}). (In Appendix[\ref{apphll}], we have shown that the relation is true for an arbitrary pair of bulk and surface term provided the surface term is homogeneous in its dynamical degrees of freedom). This suggests that holography has its deep roots in the very nature of the \ll\  itself (and hence applies to \ee as a special case). 

Our thermodynamic interpretation of the holographic relation also carries over to the \LL\ models in a straightforward manner. We can now consider the bulk and boundary terms of the action as providing the energy and entropy of the system and the holographic relation as providing the means for obtaining $S(E)$ from $E(S)$ and vice-versa.

For the Lagrangian written in two different ways (as in \eq{llsp} and \eq{llsum}) let us consider the difference between the two expressions each of which leads to the entropy when evaluated on the horizon. Using the surface term in \eq{llsum}, we define 
\begin{equation}
S_1 = m L^{(1)}_{sur} =  2m \partial_\beta\left[\sqrt{-g}Q_a^{\phantom{a}b\beta d}\Gamma^a_{bd}\right]
\end{equation}
 and using \eq{llsp},
\begin{equation}
S_2 = L^{(2)}_{sur} = 2\partial_\beta(\sqrt{-g}P_a^{\ b\beta 0}\Gamma^a_{0b})
\end{equation}
Both $S_1$ and $S_2$ terms give the Wald entropy when integrated over a horizon. This is because their difference 
\begin{eqnarray}
S_1 - S_2 &=& 2m \partial_\beta \left( \sqrt{-g} Q_a^{\ b\beta \alpha} \Gamma^a_{b\alpha} \right)
\end{eqnarray} 
vanishes on a spherically symmetric horizon as only the $Q^{0110}$ component contributes on the horizon (see Section IV.B of ref.\cite{ayantp}). The direct difference between the two surface terms $\Delta L_{sur}$ and the difference between the corresponding two bulk terms $\Delta L_{bulk}$ follow the holographic relation Eqn.[\ref{mainresult}] due to the linearity of the relation.

\section{Discussion and Conclusions}
The conventional approach to gravity begins from principle of equivalence and principle of general covariance. These are sufficient to argue that effects of gravity on matter can be given a geometrical interpretation in terms of a nontrivial metric. The (minimal) coupling of gravity to matter can be obtained by demanding the validity of special relativistic laws in local inertial frames and general covariance. Unfortunately, there is no such elegant principle to write down an action functional and obtain the dynamics of gravity. One can construct a large class of generally covariant action functionals of which the Einstein-Hilbert action is the simplest in four dimensions. 

The alternative paradigm considers gravity to be an emergent phenomenon \cite{rop,others} and obtains the field equations from thermodynamic considerations (see e.g., \cite{reviews2}). Since the same field equations are obtained by both the procedures, it seems reasonable to believe that the structure of action functionals used in the conventional approach must contain certain clues to the fact that gravity has an emergent description.

This expectation turns out to be true. Unlike all other theories in physics (in particular non-abelian gauge theories which are in some sense closest to gravity), the gravitational action --- determined by the symmetries of the theory --- has a nontrivial surface term. An investigation of the surface term shows that: (i) it has a deep connection with the entropy of horizons that arise in  certain on-shell solutions of the theory \cite{holo} and (ii) it has a holographic relation to the bulk term \cite{grf2002}. What is more, the existence of the surface term and its properties seems to be generic \cite{ayantp} and arises in a wide class of theories more general than Einstein gravity. 

In the conventional approach,  no intuitive explanation is available for these features;
but in the thermodynamic paradigm, one can interpret the bulk term as energy and the surface term as entropy. (In fact, one can argue that \cite{reviews2} the action functional in any reasonable theory of gravity should contain  a surface term). The holographic relationship between the bulk and surface term then acquires a thermodynamic interpretation and is analogous to the usual Legendre-like transformations in thermodynamics allowing one to construct the energy functional from entropy functional and vice-versa. Some explicit constructions along these lines have already been done, demonstrating the utility of the holographic relation \cite{use}.

The extra feature not present in the conventional thermodynamics is the dimensional reduction that occurs in any holographic relation. Note that the holographic relation itself can be stated as a relation between Lagrangian densities as in \eq{holoeh1}, \eq{holoeh} or \eq{hololl} etc. But on integrating the Lagrangian over a region $\mathcal{V}$ to obtain the action, the $L_{sur}$ contributes  a surface term in $\partial\mathcal{V}$. Thus
 the dynamics of gravity, expressed through the field equations in the $D$ dimensional space (``bulk" $\mathcal{V}$) is equally well encoded in a functional expressed in  the $D-1$ dimensional space (``boundary" $\partial \mathcal{V}$). The fundamental reason for this is the existence of horizons in gravitational theories and the need to encode information blocked by the horizons on its surface.

Our analysis has also revealed the relationship between Noether charge density and the surface term in the action for static geometries. One can, in fact, start from Noether charge density, construct the entropy functional and then determine the bulk term of the action though holographic condition. Such an approach requires some careful considerations of uniqueness which were sorted out for the case of Einstein gravity in previous work \cite{grf2002,holo}. We hope to address corresponding issues in the case of \LL\ models in a future work. Finally it would be interesting to extend the study to cover non-static spacetimes, which possibly might require extension of the ideas to situations away from thermodynamic equilibrium.

\section*{Acknowledgements}
SK is supported by a Fellowship from the Council of Scientific and Industrial Research (CSIR), India.

\appendix

\section{}

\subsection{Proof of \eq{ehholo} in Einstein's gravity by direct calculation}\label{apphee}
We first expand out the Euler derivative and write
\begin{eqnarray}
&&\partial_i\left[ g_{ab}\frac{\delta f}{\delta(\partial_ig_{ab})}+\partial_jg_{ab}\frac{\partial f}{\partial(\partial_i \partial_j g_{ab})} \right] \nonumber \\
&=& \partial_i\biggl[ g_{ab}\frac{\partial f}{\partial(\partial_ig_{ab})}-g_{ab}\partial_h\frac{\partial f}{\partial(\partial_h\partial_ig_{ab})} \nonumber \\
\qquad && +\partial_jg_{ab}\frac{\partial f}{\partial(\partial_i \partial_j g_{ab})} \biggr] \nonumber \\
\label{dircal}
\end{eqnarray}
We prove \eq{ehholo} in two parts. For the first part we find the following three quantities
\begin{eqnarray}
g_{np}\frac{\partial L}{\partial (\partial_m g_{np})} &=& 2\sqrt{-g}[Q_a^{\phantom{a}bac}\Gamma^m_{bc}-2Q^{nbmd}\Gamma_{ndb}]\nonumber\\
-g_{np}\partial_s\frac{\partial L}{\partial (\partial_s\partial_m g_{np})} &=& -2\sqrt{-g}[Q_a^{\phantom{a}bac}\Gamma^m_{bc}-Q^{nbmd}\Gamma_{ndb}] \nonumber \\
\partial_mg_{np}\partial_s\frac{\partial L}{\partial (\partial_s\partial_m g_{np})} &=& 2\sqrt{-g}Q^{nbmd}\Gamma_{ndb} \nonumber
\end{eqnarray}
where $L=L_{EH}\sqrt{-g}$ and $Q_a^{\phantom{a}bcd}=\frac{1}{2}(\delta^c_ag^{bd}-\delta^d_ag^{bc})$
Adding them we get 
\begin{equation}
\partial_i\left[ g_{ab}\frac{\delta L}{\delta(\partial_ig_{ab})}+\partial_jg_{ab}\frac{\partial L}{\partial(\partial_i \partial_j g_{ab})} \right]=0
\label{fpart}
\end{equation}
For the second part, we expand $L_{sur}=2\sqrt{-g}R^0_0 $ as
\begin{eqnarray}
L_{sur} 
&=& \sqrt{-g}\partial_bg_{a0}\biggl[ g^{b0}\partial_cg^{ac}+g^{ac}\partial_cg^{b0} \nonumber \\
&& -g^{a0}\partial_cg^{bc}-g^{bc}\partial_cg^{a0}\biggr] \nonumber \\
&& + \sqrt{-g}\left[ 2Q^{abc0}\partial_c\partial_bg_{a0}+Q^{abc0}g^{pq}\partial_bg_{a0}\partial_cg_{pq}\right] \nonumber
\end{eqnarray}
We calculate following quantities step by step \\
First, we have:
\begin{eqnarray}
&&\frac{\partial L_{sur}}{\partial(\partial_h \partial_k g_{ij})}=2\sqrt{-g}Q^{ikh0}\delta^j_0 \nonumber \\ \nonumber \\
\end{eqnarray}
Next,
\begin{eqnarray}
&&g_{ij}\partial_h\left( \frac{\partial L_{sur}}{\partial(\partial_h \partial_k g_{ij})} \right) \nonumber \\ \nonumber \\
&& = \sqrt{-g}g^{pq}Q^{ikh0}g_{i0}\partial_hg_{pq} \nonumber \\ \nonumber \\
&&+ \sqrt{-g} \left[ g^{k0}g_{i0}\partial_hg^{ih}-g^{kh}g_{i0}\partial_hg^{i0} + \partial_0g^{k0}-\delta^0_0\partial_hg^{kh}\right] \nonumber \\ \nonumber \\
\end{eqnarray}
The third relation we need is:
\begin{eqnarray}
&&\partial_hg_{ij}\frac{\partial L_{sur}}{\partial(\partial_k \partial_h g_{ij})}=2\sqrt{-g}Q^{ihk0}\partial_hg_{i0} \nonumber \\ \nonumber \\
\end{eqnarray}
Finally, we also have:
\begin{eqnarray}
&& g_{ij}\frac{\partial L_{sur}}{\partial(\partial_kg_{ij})}  \nonumber \\ \nonumber \\
&& = \sqrt{-g}g^{pq}Q^{ikh0}g_{i0}\partial_hg_{pq} \nonumber \\ \nonumber \\
&&+ \sqrt{-g} \left[ g^{k0}g_{i0}\partial_hg^{ih}-g^{kh}g_{i0}\partial_hg^{i0} + \partial_0g^{k0}-\delta^0_0\partial_hg^{kh}\right] \nonumber 
\end{eqnarray}
From these we find that
\begin{equation}
g_{ij}\frac{\delta L_{sur}}{\delta(\partial_kg_{ij})}=0
\end{equation}
Hence,
\begin{eqnarray}
&&\partial_i\left[ g_{ab}\frac{\delta L_{sur}}{\delta(\partial_ig_{ab})}+\partial_jg_{ab}\frac{\partial L_{sur}}{\partial(\partial_i \partial_j g_{ab})} \right] \nonumber \\
&&=\partial_i\left[\partial_jg_{ab}\frac{\partial L_{sur}}{\partial(\partial_i \partial_j g_{ab})} \right] \nonumber \\
&&=\partial_k\left( \sqrt{-g}2Q^{abk0}\partial_bg_{a0} \right) = L_{sur}
\label{spart}
\end{eqnarray}
Subtracting \ref{fpart} from \ref{spart}, we get \eq{ehholo} for $D=4$.

\subsection{Holography of Lanczos-Lovelock Lagrangian} \label{apphll}
We derive a general result which shows that there always exists a holographic relationship as defined by \eq{hololl} between any pair of bulk and surface terms of the Lanczos-Lovelock Lagrangian when the surface term is homogeneous in its variables as described below.

Let the Lanczos-Lovelock Lagrangian $L=\sqrt{-g}L_{(m)}$ in $D$ dimensions be written as a sum of bulk term $L_1$ and a total divergence term $L_2$ giving $L=L_1+L_2$. 
Let the total divergence term $L_2$ in the Lagrangian be such that it satisfies the following homogeneity condition:
\begin{itemize}
\item When the surface term $L_2$ is expanded out and written in terms of $g_{ab}, \partial_cg_{ab},\partial_d \partial_c g_{ab} $, a generic term in the expansion will have the form  $(g_{**})^x(\partial_* \partial_*g_{**})^k(\partial_* g_{**})^y$ for some indices $x,y,k$. We assume that all the terms are  homogeneous in degree $p$; that is,
\begin{equation}
x+y+k=p
\label{codn}
\end{equation}

\end{itemize}
We can then show that the Lanczos-Lovelock Lagrangian $L=\sqrt{-g}L_{(m)}$ in $D$ dimensions, with this decomposition,  is holographic in the sense that:
\begin{equation}
[D/2+p]L_2=-\partial_i\left[ g_{ab}\frac{\delta L_1}{\delta(\partial_ig_{ab})}+\partial_jg_{ab}\frac{\partial L_1}{\partial(\partial_i \partial_j g_{ab})} \right] 
\label{result}
\end{equation}
where $L_1$ is \textit{defined} through $L_1\equiv L-L_2$ and differentiation indicated by $\delta$ is the Euler derivative. The result in Eqn.[\ref{result}] is a generalization of the holographic relation proved in \cite{ayantp} for the Lanczos-Lovelock Lagrangian written in terms of a $Q\Gamma \Gamma$ bulk term and a $\nabla(Q\Gamma)$ surface term and uses the same technique. 

The proof of \eq{result} is as follows. Given a $L_2$ satisfying the homogeneity condition, we first define 
\begin{equation}
L_1=\sqrt{-g}L_{(m)}-L_2=L-L_2
\end{equation}
 (This can be simplified and written in a compact form depending on the surface term chosen; however for this proof the given form is useful.)
Consider the quantity 
\begin{equation}
\partial_i\left[ g_{ab}\frac{\delta f}{\delta(\partial_ig_{ab})}+\partial_jg_{ab}\frac{\partial f}{\partial(\partial_i \partial_j g_{ab})} \right]
\end{equation}  
One can show, after some manipulations, that
\begin{eqnarray}
 &&\partial_i\left[ g_{ab}\frac{\delta f}{\delta(\partial_ig_{ab})}+\partial_jg_{ab}\frac{\partial f}{\partial(\partial_i \partial_j g_{ab})} \right] \nonumber \\ 
& = & \left\{  g_{ab} \frac{\partial f}{\partial g_{ab}}+(\partial_ig_{ab})\frac{\partial f}{\partial(\partial_i g_{ab})}+(\partial_i\partial_jg_{ab})\frac{\partial f}{\partial(\partial_i\partial_j g_{ab})} \right\} \nonumber \\
 && -g_{ab}\frac{\delta f}{\delta g_{ab}}
\label{main}
\end{eqnarray}
Since $L_2$ is a divergence, its Euler derivative identically vanishes
\begin{equation}
\frac{\delta L_2}{\delta g_{ab}}=0
\label{E1}
\end{equation}
The Euler derivative say $M^{ab}[L]$ of the \LL\ Lagrangian $L$ satisfies the property that its trace is proportional to the Lagrangian itself 
\begin{eqnarray}
g_{ab}\frac{\delta L}{\delta g_{ab}} && =g_{ab}M^{ab}[L] = -\sqrt{-g}g_{ab}\e^{ab}[L] \nonumber \\
&&=[D/2-m]\sqrt{-g}L_m
\label{E2}
\end{eqnarray}
We will now prove  two preliminary results using the homogeneity condition of $L_2$ and the natural homogeneity of $L$. We show that:
\begin{eqnarray}
&& g_{ab}\frac{\partial L}{\partial g_{ab}} + (\partial_ig_{ab})\frac{\partial L}{\partial(\partial_i g_{ab})}+(\partial_i\partial_jg_{ab})\frac{\partial L}{\partial(\partial_i\partial_j g_{ab})} \nonumber \\
&&=(D/2-m)L \nonumber \\
\label{11}
\end{eqnarray}
and
\begin{eqnarray}
&& g_{ab}\frac{\partial L_2}{\partial g_{ab}} + (\partial_ig_{ab})\frac{\partial L_2}{\partial(\partial_i g_{ab})}+(\partial_i\partial_jg_{ab})\frac{\partial L_2}{\partial(\partial_i\partial_j g_{ab})} \nonumber \\
&&=(D/2+p)L_2 \nonumber \\
\label{l2}
\end{eqnarray}
To prove these relations, consider any generic term $f^{(k)}$ (here $k$ is a label ) which arises when $f$ is expanded in terms of $g_{ab}, \partial_cg_{ab},\partial_d \partial_c g_{ab}$, where $k$ is the degree of $\partial_d \partial_c g_{ab})$. Here $f$ is a dummy scalar and we will later put $f=L$ or $f=L_2$ according to our need. Let $x$ and $y$ be the degree of $g_{ab}$ and $ \partial_cg_{ab}$ in the term $f^{(k)}$. Hence by definition,
\begin{eqnarray}
g_{ab}\frac{\partial f^{(k)}}{\partial g_{ab}}=(D/2+x)L^{(k)}  ;&&  (\partial_ig_{ab})\frac{\partial f^{(k)}}{\partial(\partial_i g_{ab})}=yf^{(k)}; \nonumber \\
(\partial_i\partial_j g_{ab})\frac{\partial f^{(k)}}{\partial(\partial_i\partial_j g_{ab})}&=&kf^{(k)}
\label{sca}
\end{eqnarray}
 The $D/2$ factor arises due to $\sqrt{-g}$. Adding the three, we get
\begin{eqnarray}
&&\left\{ g_{ab}\frac{\partial f^{(k)}}{\partial g_{ab}}+(\partial_ig_{ab})\frac{\partial f^{(k)}}{\partial(\partial_i g_{ab})}+(\partial_i\partial_jg_{ab})\frac{\partial f^{(k)}}{\partial(\partial_i\partial_j g_{ab})} \right\} \nonumber \\
&&=(D/2+(x+y+k))f^{(k)}
\end{eqnarray}
When $f^{(k)}=L_2^{(k)}$, the homogeneity condition on $L_2$ tells us $x+y+k=p$, which is independent of the $k$th term and hence true for any generic term in $L_2$. Hence the above expression is valid for the $L_2$, and leads us to Eqn.[\ref{l2}].\\

The \LL\ Lagrangian $L=\sqrt{-g}L_{(m)}$ also satisfies the homogeneity condition naturally with the degree $p=-m$. To see this, we note that the scalar $L$ is made up from the metric, its first and second derivatives; hence the upper indices must be equal to the number of lower indices in any generic term of $L$ giving us the relation $2(-x)=3y+4k$. This fixes the number of $g_{ab}$ in terms of $\partial_ig_{ab}$ and $\partial_i\partial_jg_{ab}$. Since the \LL\ Lagrangian $L$ is made up of a product of $m$ curvature  tensors $R \sim \partial^2 g + (\partial g)^2$, the $L^{(k)}$ term will have $k$ factors of $\partial^2 g$ and $(m-k)$ factors of $(\partial g)^2$, that is $y = 2(m-k)$ and hence $(-x)=3m-k$. This gives $x+y+k=-m$ which is again independent of the $k$th term and hence leads us \eq{11}.

Using \eq{E2} and \eq{11} in \eq{main}, we get
\begin{equation}
\partial_i\left[ g_{ab}\frac{\delta L}{\delta(\partial_ig_{ab})}+\partial_jg_{ab}\frac{\partial L}{\partial(\partial_i \partial_j g_{ab})} \right]=0
\label{result1}
\end{equation}
Using \eq{E1} and \eq{l2} in \eq{main}, we get
\begin{equation}
[D/2+p]L_2=\partial_i\left[ g_{ab}\frac{\delta L_2}{\delta(\partial_ig_{ab})}+\partial_jg_{ab}\frac{\partial L_2}{\partial(\partial_i \partial_j g_{ab})} \right]
\label{result2}
\end{equation}
Writing $L_2$ in the RHS of the above equation as $L_2=L-L_1$ and using \eq{result1},we get the holographic relation of \eq{result}
\begin{equation}
[D/2+p]L_2=-\partial_i\left[ g_{ab}\frac{\delta L_1}{\delta(\partial_ig_{ab})}+\partial_jg_{ab}\frac{\partial L_1}{\partial(\partial_i \partial_j g_{ab})} \right]
\label{final}
\end{equation}
One can now check that the surface terms we discussed satisfy the homogeneity condition.
(i) Consider the form of $L_m$ where $L_{sur}= \sqrt{-g}\nabla_a J^{0a}$ in \eq{llsp}.
Expanding $\sqrt{-g}\nabla_a J^{0a}$ out in terms of metric and its derivatives, one can easily see that it satisfies the homogeneity condition with $p=-m$.
(ii)  $L_{sur}=2\partial_c \left[ \sqrt{-g}Q_a^{bcd}\Gamma^a_{bd} \right]$ in \eq{llsum} is homogeneous with degree $p=-m$. Hence the holographic relationship follows.

\section{}
\subsection{Variation of the \ll}\label{appllvary}
The variation of the quantity $L\sqrt{-g}$ where $L$ is the \ll\ can be expressed as
\begin{eqnarray}
\delta \left(L\sqrt{-g}\right) & = & \left(\frac{\partial L\sqrt{-g}}{\partial g^{ab}}\right)\, \delta g^{ab} + \left(\frac{\partial L\sqrt{-g}}{\partial R^a_{\ bcd}} \right)\, \delta R^a_{\ bcd} \nonumber \\
& = & \left(\frac{\partial L\sqrt{-g}}{\partial g^{ab}} \right)\,\delta g^{ab} + \sqrt{-g} P_a^{\ bcd}\, \delta R^a_{\ bcd} \nonumber \\
\label{ethree}
\end{eqnarray} 
The term $P_a^{\ bcd}\, \delta R^a_{\ bcd}$ is generally covariant
and hence can be evaluated in the local inertial frame using 
\begin{eqnarray}
\delta R^a_{\ bcd}
& = & \nabla_c \left(\delta \Gamma^a_{db}\right) - \nabla_d \left(\delta \Gamma^a_{cb}\right) \nonumber \\
& = & \frac{1}{2} \nabla_c\left[
g^{ai}\left(-\nabla_i\delta g_{db} + \nabla_d\delta g_{bi} + \nabla_b\delta g_{di} \right)\right] \nonumber \\
&& - \{\textrm{term with } c\leftrightarrow d\} \nonumber \\
\end{eqnarray}
Multiplying this expression by $P_a^{\ bcd}$ the middle term $g^{ai}\nabla_d\delta g_{bi}$ will not contribute because of the anti symmetry of 
$P^{ibcd}$ in $i$ and $b$. The other two terms will contribute equally. We will get
a similar contribution from the term with $c$ and $d$ interchanged. Thus
\begin{eqnarray}
P_a^{\ bcd}\delta R^a_{\ bcd} & = & 2 \nabla_c [P^{ibcd} \nabla_b (\delta g_{di})] \nonumber \\
& = & 2 \nabla_c [P_i^{\ bcd} \delta \Gamma^i_{bd}]
\label{efive}
\end{eqnarray} 
To find $\left(\frac{\partial L\sqrt{-g}}{\partial g^{ab}} \right)$, we write
\begin{eqnarray}
\left(\frac{\partial L\sqrt{-g}}{\partial g^{ab}} \right) & = &\left ( \frac{\partial L}{\partial R^{kl}_{ij}}\frac{\partial R^{kl}_{ij}}{\partial g^{ab}} - \frac{1}{2}g_{ab}L \right) \nonumber \\
& = & \left ( P^{ij}_{kb}R^k_{\ aij}- \frac{1}{2}g_{ab}L \right) \nonumber \\
& = & \left ( P^{\ kij}_{b}R_{akij}- \frac{1}{2}g_{ab}L \right)
\end{eqnarray}
where in arriving at the first inequality we have used the fact that while differentiating $R_{ij}^{kl}=g^{lm}R^k_{\ mij}$ we should keep $R^k_{\ mij}$ fixed.
Hence we get,
\begin{eqnarray}
\delta ( L\sqrt{-g}) & = & \left(  P^{\ kij}_{b}R_{akij}- \frac{1}{2}g_{ab}L \right)\delta g_{ab} \nonumber \\
&& + \sqrt{-g}\nabla_j \left[ 2 P_i^{\ bjd} \delta \Gamma^i_{bd} ) \right] \nonumber \\
& = & \sqrt{-g} \left( \e_{ab}\delta g^{ab}+\nabla_a \delta\upsilon ^a \right)
\label{lvary}
\end{eqnarray}

\subsection{Variation of the $Q\Gamma \Gamma$ term}\label{appggvary}
From \eq{llsum}, we write
\begin{eqnarray}
L_{bulk} & = & 2\sqrt{-g}Q_a^{\ bcd}\Gamma^a_{dk}\Gamma^k_{bc} \nonumber \\
& = & \sqrt{-g}L - 2\partial_c\left[\sqrt{-g}Q_a^{\ abcd}\Gamma^a_{bd}\right]
 \end{eqnarray} 
Hence, we get 
\begin{equation}
\delta L_{bulk} = \delta ( L\sqrt{-g}) - \delta L_{sur}
\label{ggg}
\end{equation}
Using the definition of $L_{sur}$ from \eq{llsum}, we find that
\begin{eqnarray}
\delta L_{sur} = 2Q^{cd}_{ak}\partial_c \left [ \sqrt{-g}g^{bk}\delta \Gamma^a_{bd} +  \Gamma^a_{bd} \delta (\sqrt{-g}g^{bk}) \right ]
\label{one}
\end{eqnarray}
Using
\begin{eqnarray}
\delta (\sqrt{-g}g^{bk}) & = & \sqrt{-g} \left [ \delta^b_l \delta^k_m - \frac{1}{2}g^{bk}g_{lm} \right ]\delta g^{lm} \nonumber \\
& =& \sqrt{-g} B^{bk}_{lm}\delta g^{lm}
\end{eqnarray}
where the last equality defines $B^{bk}_{lm}$, we can write the second term in \eq{one} as
\begin{eqnarray}
2Q^{cd}_{ak}\partial_c \left [ \Gamma^a_{bd} \delta (\sqrt{-g}g^{bk}) \right ] & = & 2Q^{cd}_{ak}\partial_c \left [ \sqrt{-g} \Gamma^a_{bd} B^{bk}_{lm}\delta g^{lm} \right ] \nonumber \\
& \equiv & \partial_c [\sqrt{-g} M^c_{\ lm} \delta g^{lm}]
\end{eqnarray}
where we have defined the 3 index non-tensorial object
\begin{eqnarray}
M^c_{\ lm} & = & 2Q^{cd}_{ak} \Gamma^a_{bd} B^{bk}_{lm} \nonumber \\
& = & \Gamma^c_{lm} - \Gamma^d_{ld} \delta^c_m+\frac{g_{lm}}{2g}\partial_b(gg^{bc})
\label{mdef}
\end{eqnarray}
Hence,
\begin{eqnarray}
\delta L_{sur} = \partial_c [2 \sqrt{-g}g^{bk} Q^{cd}_{ak} \delta \Gamma^a_{bd} + \sqrt{-g} M^c_{\ lm} \delta g^{lm}]
\label{lsurvary}
\end{eqnarray}
Using \eq{lvary} and \eq{lsurvary} in \eq{ggg}, we get
\begin{eqnarray}
\delta L_{bulk} & = & \sqrt{-g} \e_{ab}\delta g^{ab} + (m-1)\partial_c \left [2 \sqrt{-g}g^{bk} Q^{cd}_{ak} \delta \Gamma^a_{bd} \right ] \nonumber \\
&& - \partial_c \left [ \sqrt{-g} M^c_{\ lm} \delta g^{lm} \right ]
\end{eqnarray}
Note that in the Einstein's gravity case $m=1$ the second term vanishes and we are only required to fix the metric at the boundary. However in the general case one has to fix the metric as well as the normal derivative of the metric at the boundary.

\subsection{Variation of $-2\sqrt{-g}\e^0_0$}\label{appg00vary}
We write
\begin{eqnarray}
\delta(-2\sqrt{-g}\e^0_0) = \delta ( L\sqrt{-g}) - \delta (2\sqrt{-g}{\cal R}^0_0)
\end{eqnarray}
Using the definition of $2{\cal R}^0_0$ for static spacetime in \eq{llsp}
\begin{equation}
2\sqrt{-g}{\cal R}^0_0=2m\partial_c(\sqrt{-g}Q_a^{\ bc0}\Gamma^a_{0b})
\end{equation}
we proceed in the manner similar to Appendix[\ref{appggvary}]. We then find
\begin{eqnarray}
\delta (2\sqrt{-g}{\cal R}^0_0) = m \partial_c [2 \sqrt{-g}g^{bk} Q^{c0}_{ak} \delta \Gamma^a_{b0} + \sqrt{-g} M^c_{\ lm} \delta g^{lm}] \nonumber \\
\end{eqnarray}
where now the 3 index non-tensorial object is defined as
\begin{eqnarray}
M^c_{\ lm} & = & 2Q^{c0}_{ak} \Gamma^a_{b0} B^{bk}_{lm} \nonumber
\end{eqnarray}
Note that this is different from the second equality in \eq{mdef}.
From these and \eq{lvary}, we get
\begin{eqnarray}
 \delta(-2\sqrt{-g}\e^0_0) & = & \sqrt{-g} \e_{ab}\delta g^{ab} + m\partial_c \left [2 \sqrt{-g}g^{bk} Q^{c\alpha}_{ak} \delta \Gamma^a_{b\alpha} \right ] \nonumber \\
&& - m\partial_c \left [ \sqrt{-g} M^c_{\ lm} \delta g^{lm} \right ]
\end{eqnarray}
Unlike the situation in Appendix[\ref{appggvary}], the second term in the above equation does not vanish trivially even for the Einstein's gravity which corresponds to $m=1$. However, note that the entire analysis is relevant only to in the context of static spacetimes and hence we restrict our variations to metrics which are static; that is we choose a coordinate system in which the Killing vector has the components $\xi^a=(1,\mathbf{0})$ and consider variations of the form $g^{ab}(\mathbf{x})\to g^{ab}(\mathbf{x})+\delta g^{ab}(\mathbf{x})$. Then only spatial derivatives survive on the boundaries and the variation is well-defined leading to the static Einstein equations.


\end{document}